\documentclass[pre,aps,twocolumn]{revtex4}

\usepackage{graphicx}
\usepackage{amsmath}
\usepackage{amssymb}
\usepackage{ifthen}
\usepackage{booktabs}

\usepackage{graphicx}
\usepackage{dcolumn}
\usepackage{bm}
\usepackage{longtable}
\usepackage{gensymb}

\newcommand{\bb}{\begin{equation}}
\newcommand{\ee}{\end{equation}}
\newcommand{\ba}{\begin{eqnarray*}}
\newcommand{\ea}{\end{eqnarray*}}
\newcommand{\rhor}{\rho({\bf r})}
\newcommand{\dd}{{\rm d}}
\newcommand{\rr}{{\mathbf r}}
\newcommand{\dr}{{\rm d}{\bf r}}

\begin{document}

\title{Continuous condensation in nanogrooves}

\author{Alexandr \surname{Malijevsk\'y}}
\affiliation{
{Department of Physical Chemistry, University of Chemical Technology Prague, Praha 6, 166 28, Czech Republic;}\\
{Department of Microscopic and Mesoscopic Modelling, ICPF of the Czech Academy of Sciences, 165 02 Prague 6, Czech Republic}}

\begin{abstract}
\noindent We consider condensation in a capillary groove of width $L$ and depth $D$, formed by walls that are completely wet (contact angle
$\theta=0$), which is in a contact with a gas reservoir of the chemical potential $\mu$. On a mesoscopic level, the condensation process can be
described in terms of the midpoint height $\ell$ of a meniscus formed at the liquid-gas interface. For macroscopically deep grooves ($D\to\infty$),
and in the presence of long-range (dispersion) forces, the condensation corresponds to a second order phase transition, such that $\ell\sim
(\mu_{cc}-\mu)^{-1/4}$  as $\mu\to\mu_{cc}^-$ where $\mu_{cc}$ is the chemical potential pertinent to capillary condensation in a slit pore of width
$L$. For finite values of $D$, the transition becomes rounded and the groove becomes filled with liquid at a chemical potential higher than
$\mu_{cc}$ with a difference of the order of $D^{-3}$. For sufficiently deep grooves, the meniscus growth initially follows the power-law $\ell\sim
(\mu_{cc}-\mu)^{-1/4}$ but this behaviour eventually crosses over to $\ell\sim D-(\mu-\mu_{cc})^{-1/3}$ above $\mu_{cc}$, with a gap between the two
regimes shown to be $\bar{\delta}\mu\sim D^{-3}$. Right at $\mu=\mu_{cc}$, when the groove is only partially filled with liquid, the height of the
meniscus scales as $\ell^*\sim (D^3L)^{1/4}$. Moreover, the chemical potential (or pressure) at which the groove is half-filled with liquid exhibits
a non-monotonic dependence on $D$ with a maximum at $D\approx 3L/2$ and coincides with $\mu_{cc}$ when $L\approx D$. Finally, we show that
condensation in finite grooves can be mapped on the condensation in capillary slits formed by two asymmetric (competing) walls a distance $D$ apart
with potential strengths depending on $L$. All these predictions, based on mesoscopic arguments, are confirmed by fully microscopic Rosenfeld's
density functional theory with a reasonable agreement down to surprisingly small values of both $L$ and $D$.
\end{abstract}

\maketitle

\section{Introduction}

Capillary condensation, i.e. the phenomenon whereby an undersaturated gas confined by solid walls condenses to a high density, liquidlike phase, is
perhaps the most fundamental manifestation of surface tension and finite-size effects \cite{row,evans90}. For sufficiently wide pores, the difference
in the chemical potential $\mu_{cc}$ at which the fluid condenses in an open slit pore formed of two parallel plates from the saturated chemical
potential $\mu_{\rm sat}$ is provided by the classical Kelvin equation which can be derived by combining the geometric properties of the pore with
the properties of a meniscus formed at the gas-liquid interface. According to this macroscopic viewpoint, the meniscus is of a circular cross-section
with a Laplace radius and meets both walls at Young's contact angle $\theta$. However, the Kelvin equation becomes increasingly less reliable as the
pore width $L$ is decreased towards a molecular scale, especially when the side walls are completely wet $\theta=0$. In this case, the slit walls are
covered by liquidlike layers of width $\ell_\pi$, so that the space between the walls available for the gas molecules is effectively reduced. The
importance of the presence of the wetting layers on the location of capillary condensation was firstly recognized by Derjauguin's school using the
concept of disjoining pressure \cite{der} which has been later put into a more general picture of interfacial phenomena by Evans {\it et al.}
\cite{ev_mar85,ev_mar_tar86} within the framework of density functional theory  \cite{evans79}. It follows that the location of the capillary
condensation phase transition in a slit exerting van der Waals (dispersion) forces is given by the modified Kelvin equation
 \bb
 \mu_{cc}(L)=\mu_{\rm sat}-\frac{2\gamma}{\Delta\rho(L-3\ell_\pi)} \label{kelvin}
 \ee
where $\gamma$ is the gas-liquid surface tension and $\Delta\rho=\rho_l-\rho_g$ is the difference between the particle densities of the coexisting
bulk phases. Note that the presence of the factor $3$ in the denominator does not reflect the geometric restriction of the volume available to the
gas molecules but rather the nature of the asymptotic behavior of the wall-fluid interaction.

More recently, fluid condensation in a much more experimentally realistic model of capillary grooves etched into a solid surface has attracted
substantial attention \cite{darbellay,evans_cc,tasin,mistura,hofmann,
schoen,mal_groove,parry_groove,ser,mal_13,mistura13,our_groove,monson,fan,bruschi}. Within this model, one considers an infinitely long slit of width
$L$ and depth $D$ which is in contact with the bulk gas via the open top. For macroscopically deep grooves $D\to\infty$, the recent theoretical and
experimental studies revealed that the nature of condensation in grooves differs considerably from that in open slits. Perhaps most importantly, the
transition remains first-order only when the contact angle of the groove wall is finite $\theta>0$. However, above the wetting temperature $T_w$ when
the walls are completely wet by liquid, i.e. $\theta=0$, the condensation turns to be continuous (critical). In this case, the wetting layers at the
side and bottom walls merge to form a meniscus, the mid-height $\ell$ of which increases continuously as $\mu_{cc}$ is approached from below and
eventually diverges according to the power-law:
 \bb
  \ell\sim\delta\mu^{-\beta_C}\,, \label{betac}
 \ee
 as $\delta\mu\equiv\mu_{cc}-\mu\to0^+$. The critical exponent $\beta_C$ depends on the asymptotic behaviour of the microscopic interactions and for van der
 Waals forces $\beta_C=1/4$.

In this work, we consider \emph{microscopically} deep grooves formed of walls interacting with the fluid via long range dispersion forces and we ask
what the repercussions of finiteness of $D$ are on the fluid condensation. As has been shown recently \cite{fin_groove_prl}, the condensation remains
continuous, although not critical since $\ell$ can not diverge anymore, so that the process does not exhibit any singular behaviour. However, we will
show that condensation in finitely deep grooves in the presence of long-range forces experiences some new aspects, not present in the case of
macroscopically deep groves, as a consequence of the competition between effective repulsions from the groove top and groove bottom acting on the
meniscus. To this end, we will compare analytic predictions based on a mesoscopic slab model with a microscopic fundamental measure density
functional theory (DFT) which takes the packing effects, that for the highly geometrically restricted systems are of crucial importance, accurately
into account and obeys the statistical mechanical sum rules, as opposed to some less sophisticated DFT versions.

In the remainder of this paper we will first formulate the microscopic model based on DFT and determine the external field exerted by the groove
walls (section II). In section III we revisit the slab model used previously to study the criticality of condensation in infinitely deep grooves
\cite{mal_groove} which we extend for finite $D$ and compare with the results obtained numerically from DFT. We conclude by summarizing and
discussing the main results in section IV and show some details of our calculations based on the slab model in Appendix A.

\section{Microscopic model}

In the classical density functional theory \cite{evans79}, the equilibrium density profile minimises the grand potential functional
 \bb
 \Omega[\rho]={\cal F}[\rho]+\int\dd\rr\rhor[V(\rr)-\mu]\,,\label{om}
 \ee
where $\mu$ is the chemical potential, and $V(\rr)$ is the external potential. The intrinsic free energy functional ${\cal F}[\rho]$ can be separated
into an exact ideal gas contribution and an excess part:
  \bb
  {\cal F}[\rho]=\beta^{-1}\int\dr\rho(\rr)\left[\ln(\rhor\Lambda^3)-1\right]+{\cal F}_{\rm ex}[\rho]\,,
  \ee
where $\Lambda$ is the thermal de Broglie wavelength and $\beta=1/k_BT$ is the inverse temperature. As is common in the modern DFT approaches, the
excess part is modelled as a sum of hard-sphere and attractive contributions where the latter is treated in a simple mean-field fashion:
  \bb
  {\cal F}_{\rm ex}[\rho]={\cal F}_{\rm hs}[\rho]+\frac{1}{2}\int\int\dd\rr\dd\rr'\rhor\rho(\rr')u_{\rm a}(|\rr-\rr'|)\,, \label{f}
  \ee
where  $u_{\rm a}(r)$ is the attractive part of the fluid-fluid interaction potential.

The fluid atoms are assumed to interact with one another via the truncated (i.e., short-ranged) and non-shifted Lennard-Jones-like potential
 \bb
 u_{\rm a}(r)=\left\{\begin{array}{cc}
 0\,;&r<\sigma\,,\\
-4\varepsilon\left(\frac{\sigma}{r}\right)^6\,;& \sigma<r<r_c\,,\\
0\,;&r>r_c\,.
\end{array}\right.\label{ua}
 \ee
which is cut-off at $r_c=2.5\,\sigma$, where $\sigma$ is the hard-sphere diameter.

The hard-sphere part of the excess free energy is approximated using the (original) Rosenfeld fundamental measure functional \cite{ros},
 \bb
{\cal F}_{\rm hs}[\rho]=\frac{1}{\beta}\int\dd\rr\,\Phi(\{n_\alpha\})\,,\label{fmt}
 \ee
which accurately takes into account the short-range correlations between the fluid particles.

We assume that the confining walls are formed by atoms distributed uniformly with a density $\rho_w$ and interact with the fluid particles via
Lennard-Jones $12$-$6$ potential $\phi_w(r)$ with the parameters $\varepsilon_w$ and $\sigma$:
 \bb
 \phi_w(r)=4\varepsilon_w\left[\left(\frac{\sigma}{r}\right)^{12}-\left(\frac{\sigma}{r}\right)^{6}\right]\,.
 \ee
The wall potential containing a single groove of depth $D$ and width $L$, which is located at $0<x<L$ and $0<z<D$, can be expressed as follows:
 \bb
 V(x,z)=V_{9-3}(z)+V_D(x,z)+V_D(L-x,z)\,, \label{single}
 \ee
except for the region  corresponding to the domain of the wall in which case $V(x,z)=\infty$ meaning the wall is impenetrable. We assume that the
groove is macroscopically long and the system is thus translation invariant along the $y$-axis. Here, $V_{9-3}(z)$ is the well known $9$-$3$
Lennard-Jones potential due to a planar wall placed at $z<0$:
 \bb
 V_{9-3}(z)=4\pi\varepsilon_w\rho_w\sigma^3\left[\frac{1}{45}\left(\frac{\sigma}{z}\right)^9-\frac{1}{6}\left(\frac{\sigma}{z}\right)^3\right]\label{vpi}\,.
 \ee

The potential $V_D(x,z)$, corresponding to a semi-infinite slab of height $D$ placed at $x<0$ and $0<z<D$, is determined by evaluating the triple
integral:
 \begin{eqnarray}
V_D(x,z)&=&\rho_w\int_{-\infty}^0\dd x'\int_{-\infty}^\infty\dd y'\int_0^{D}\dd z'\nonumber\\
&&\times\phi_w\left(\sqrt{(x-x')^2+y'^2+(z-z')^2}\right)\\
 &=&\alpha_6[\psi_6(x,z)+\psi_6(x,D-z)]\nonumber\\
 &+&\alpha_{12}[\psi_{12}(x,z)+\psi_{12}(x,D-z)]\nonumber
 \end{eqnarray}
 where
 \bb
\psi_6(x,z)=\frac{2x^4+x^2z^2+2z^4}{2x^3z^3\sqrt{x^2+z^2}}-\frac{1}{z^3}
 \ee
 and
  \begin{widetext}
 \bb
\psi_{12}(x,z)=\frac{1}{128}{\frac {128\,{x}^{16}+448\,{x}^{14}{z}^{2}+560\,{x}^{12}{z}^{4}+280\,{
x}^{10}{z}^{6}+35\,{x}^{8}{z}^{8}+280\,{x}^{6}{z}^{10}+560\,{x}^{4}{z} ^{12}+448\,{z}^{14}{x}^{2}+128\,{z}^{16}}{{z}^{9}{x}^{9} \left( {x}^{2
}+{z}^{2} \right) ^{7/2}}} -\frac{1}{z^9}\ee
 \end{widetext}
 with $\alpha_6=-\pi\varepsilon_w\rho_w\sigma^6/3$ and $\alpha_{12}=2\pi\varepsilon_w\rho_w\sigma^{12}/45$.


Minimization of (\ref{om}) leads to the Euler-Lagrange equation
 \bb
 V(\rr)+\frac{\delta{\cal F}_{\rm hs}[\rho]}{\delta\rho(\rr)}+\int\dd\rr'\rho(\rr')u_{\rm a}(|\rr-\rr'|)=\mu\,,\label{el}
 \ee
which can be solved iteratively on an appropriately discretized two dimensional grid $(0,x_m)\times(0,z_m)$. The system size is determined by the
values $x_m>L$ and $z_m>D$ that are chosen large enough to justify the following boundary conditions that we impose: $\rho(x,z_m)=\rho_b$ and
$\rho(0,z)=\rho(x_m,z)=\rho_\pi(z)$ for $z>D$ where $\rho_b$ is the reservoir gas density and $\rho_\pi(z)$ is a 1D density profile of the model
fluid near a planar wall.

Prior applying our microscopic model to study condensation in capillary grooves, we investigate the bulk phase behaviour of the model fluid by
setting $V(\rr)=0$ in Eq.~(\ref{el}); in particular we obtain that the bulk critical temperature corresponds to $k_BT_c/\varepsilon=1.414$.
Furthermore, by setting $V(\rr)=V_{9-3}(z)$ with a parameter $\rho_w\varepsilon_w=1\cdot\varepsilon\sigma^{-3}$ we find that the wall-fluid system
exhibits a first-order wetting phase transition at temperature $T_w=0.8\,T_c$. Finally, a slit model formed by a pair of parallel walls with the
total potential $V(\rr)=V_{9-3}(z)+V_{9-3}(L-z)$ for the slit width $L=20\,\sigma$ and temperature $T=1.15\,T_w$ (to be considered later for the
groove models) experiences capillary condensation at the chemical potential $\mu_{cc}=-4.018\,\varepsilon$. We note that although the external
potentials and thus the resulting equilibrium density profiles vary only in one dimension in these systems, we treat them in the same way as the
groove model, i.e. we determine $\rhor=\rho(x,z)$ for the sake of numerical consistency.

\section{Results}

\begin{figure}
\centerline{\includegraphics[width=6cm]{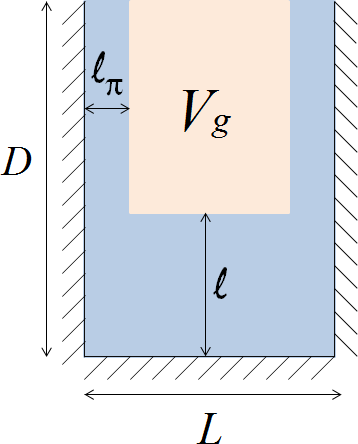}} \caption{Illustration of the slab model applied for a groove of depth $D$ and width $L$. The adsorbing
wetting layers are of the width $\ell_\pi$ and the height of the ``meniscus'' is $\ell$. The remaining volume $V_g$ is occupied by a gas-like
phase.}\label{slab}
\end{figure}

In this section we present the DFT results of the condensation in grooves of finite depth $D$ and width $L$ formed of completely wet walls and
compare with predictions based on a slab model \cite{dietrich}. The slab model has been used previously \cite{mal_groove, our_groove} to analyze the
critical behaviour of condensation and evaporation in infinitely deep grooves and it is straightforward to extend the analysis for grooves of finite
depths. Within the model one assumes that the one-body fluid density $\rhor=\rho(x,z)$ adopts only three values: i) $\rhor=\rho_l$, the liquid
density at bulk two-phase coexistence, if the fluid occupies the volume near the side and bottom walls as described in Fig.~\ref{slab}; here, the
meniscus is approximated by a flat interface of height $\ell$ above the groove bottom, while the width of the wetting layers at each of the side
walls is $\ell_\pi$; ii) $\rhor=\rho_g$, the vapour density at bulk two-phase coexistence, if the fluid occupies the region of the cross-section area
$V_g=(L-2\ell_\pi)(D-\ell)$; and (iii) $\rhor=0$ otherwise, since the groove walls are impenetrable. As in our microscopic model described in the
previous section we consider grooves that are macroscopically long, so that we assume that the system is translation invariant along the $y$-axis
normal to the sketch of Fig.~\ref{slab}.

The groove is in contact with a gas reservoir of pressure $p$, so that the adsorbed liquid phase which is metastable in bulk must have a lower
pressure $p^\dagger$. For this model, the excess (relative to the groove completely filled with liquid) grand potential per unit length is given by
 \begin{eqnarray}
 \omega^{\rm ex}(\tilde{\ell})&=&(p^\dagger-p)(L-2{\ell}_\pi)(D-\tilde{\ell})\nonumber\\
                      &&+\gamma[2(D-\tilde{\ell})+(L-2{\ell}_\pi]\nonumber\\
                       &&-\Delta\rho\int_{\tilde{\ell}}^D\dd z\int_{\ell_{\pi}}^{L-\ell_{\pi}}\dd x V(x,z) \label{slab_model}
 \end{eqnarray}
where $V(\rr)=V(x,z)$ is the external potential of the groove walls as given by Eq.~(\ref{single}). Approximating the pressure difference
$p-p^\dagger\approx\Delta\rho(\mu_{\rm sat}-\mu)$, obtained by Taylor expansion of $p(\mu)$ around $\mu_{\rm sat}$ to first order, the equilibrium
mean height of the liquid slab $\ell$ is determined by minimizing $\omega^{\rm ex}(\tilde{\ell})$ which leads to
 \begin{eqnarray}
 \mu(\ell)\approx \mu_{cc}(L)+ a\left[\frac{2}{(D-\ell)^3}-\frac{9}{8}\frac{L}{\ell^4}\right]\,. \label{EL_slab}
 \end{eqnarray}
Here, we made use of Eq.~(\ref{kelvin}) and introduced $a=\pi\varepsilon_w\rho_w\sigma^6/3$ related to the Hamaker constant. The corresponding
thermodynamic excess grand potential can be expressed in the form of the series:
\begin{eqnarray}
\frac{\omega^{\rm ex}}{(L-2\ell_\pi)} &=&\delta \mu\Delta\rho\ell+\frac{\pi L\varepsilon_w}{8\ell^3}+\cdots\nonumber\\
&&+\frac{\pi\varepsilon_w}{3(D-\ell)^2}+\cdots\label{omegaex}
\end{eqnarray}
where the ellipses denote the higher order terms in $1/\ell$ and $1/(D-\ell)$ and where we ignored irrelevant terms not dependent on $\ell$. In
deriving Eq.~(\ref{EL_slab}), all the contributions beyond the leading order terms in $1/\ell$ and $1/(D-\ell)$ have been neglected which suggests
that (\ref{EL_slab}) is only reliable for i) sufficiently deep grooves and ii) near $\mu_{cc}$; it is only in this case when both $\ell$ and $D-\ell$
are supposed to be large (compared, e.g., to $\sigma$).

\begin{figure}[h]
\centerline{\includegraphics[width=8cm]{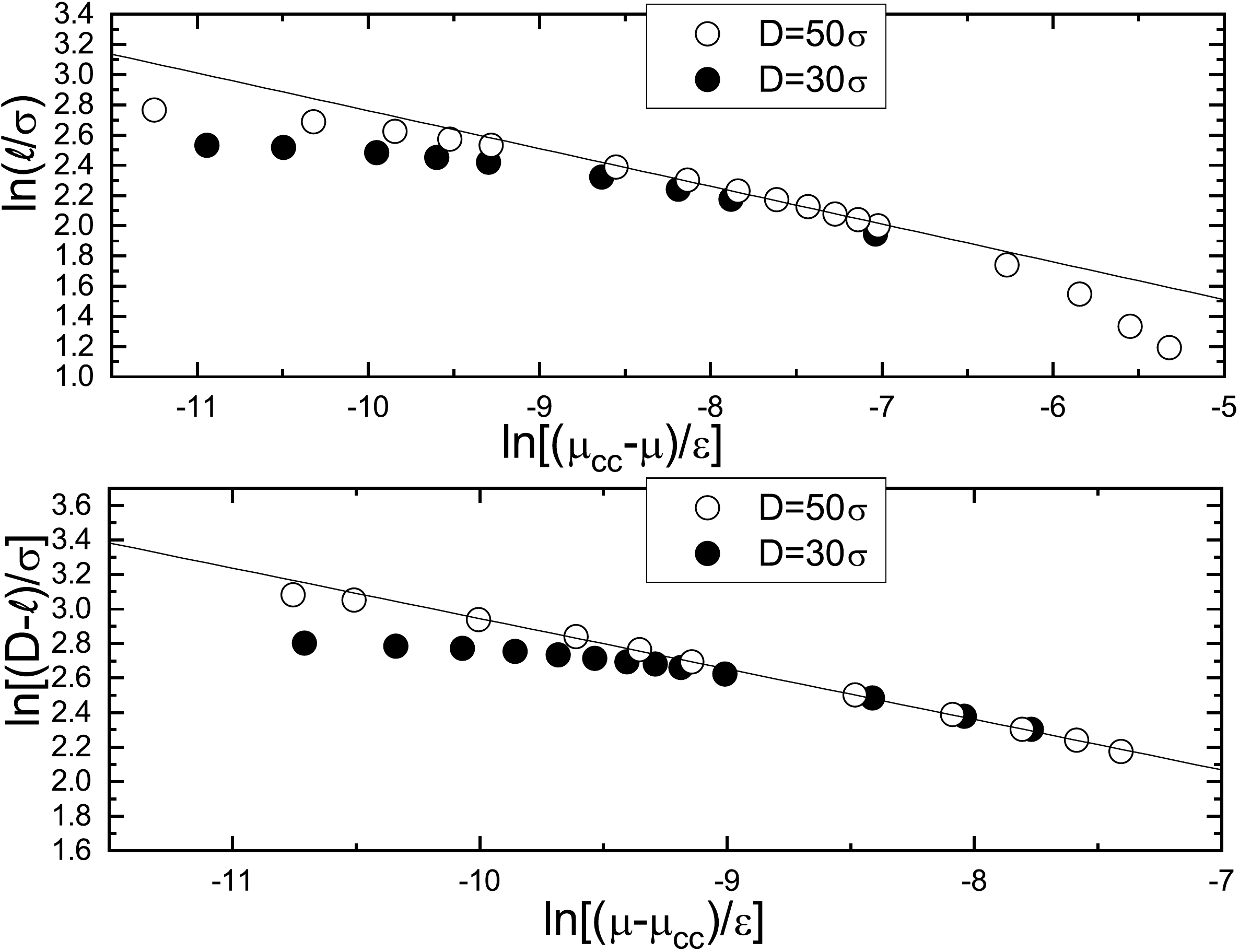}}
 \caption{A log-log plot for condensation in grooves of depth $D=30\,\sigma$ and $D=50\,\sigma$, each of width $L=20\,\sigma$,
 above the wetting temperature $T=1.15\,T_w$. The upper panel describes the regime $\mu<\mu_{cc}$ and the straight line has a slope $-1/4$,
 while the lower panel describes the regime $\mu>\mu_{cc}$ and the straight line has a slope $-1/3$.}\label{loglog}
\end{figure}

For infinitely (macroscopically) deep grooves, $D\to\infty$, Eq.~(\ref{EL_slab}) immediately reproduces the value of the critical exponent
$\beta_C=1/4$, corresponding to a divergence of $\ell$ as $\mu\to\mu_{cc}^-$; in this case the terms ${\cal{O}}((D-\ell)^{-3})$ can be neglected.
Moreover, it also implies a divergence of $D-\ell$ as $\mu_{cc}$ is approached from above pertinent to evaporation of capillary liquid, characterized
by a critical exponent $\beta_E=1/3$, in which case the terms ${\cal{O}}(\ell^{-4})$ can be neglected. For grooves of finite depths both
${\cal{O}}((D-\ell)^{-3})$ and ${\cal{O}}(\ell^{-4})$ are relevant and represent competing repulsions from the groove bottom and groove top acting
effectively on the meniscus of height $\ell$. Hence, condensation in grooves of finite depths undergoes two regimes: $\mu<\mu_{cc}$ in which case the
term $L/\ell^4$ in Eq.~(\ref{EL_slab}) representing repulsion from the groove bottom is dominating and $\mu>\mu_{cc}$ in which case the repulsion
from the groove top $\propto 1/(D-\ell)^3$ prevails. The repulsion from the groove top implies that grooves of finite depths become largely filled
with liquid at a chemical potential $\mu_f$ which is higher than $\mu_{cc}$ as for infinitely deep grooves. Introducing a small parameter $\alpha$ by
defining a meniscus height of a filled groove $\ell=(1-\alpha)D$ where $\alpha\ll1$ and substituting to Eq.~(\ref{EL_slab}), we find that
$\mu_f\approx\mu_{cc}+2a/(\alpha D)^3$.

The impact of the finite depth of grooves has been inspected by considering two systems with $D=50\,\sigma$ and $D=30\,\sigma$, both of width
$L=20\,\sigma$, at temperature $T=1.15\,T_w$. The DFT results are displayed in Fig.~\ref{loglog} where we show the dependence of $\ell(\mu)$ below
and above $\mu_{cc}$ as a log-log plot. We observe a crossover from the $\ell\propto\delta\mu^{-1/4}$ behaviour to the
$D-\ell\propto|\delta\mu|^{-1/3}$ dependance, such that the two regimes are separated by a gap of the order of $\bar{\delta}\mu$. In Appendix A we
show that
 \bb
  \bar{\delta}\mu=\frac{a}{D^3}\,,\label{bar_delta}
 \ee
as follows from our slab model. This result can be quantitatively verified by substituting the values corresponding to the DFT model at the
considered temperature, which yields $a\approx0.3$. It then follows that $\ln(\bar{\delta}\mu)\approx-9$ for $D=30\,\sigma$ and
$\ln(\bar{\delta}\mu)\approx-10.5$ for $D=50\,\sigma$, with a fairly good agreement with the DFT results.

\begin{figure}[h]
\includegraphics[width=4cm]{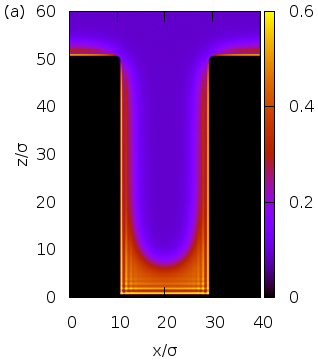} \includegraphics[width=4cm]{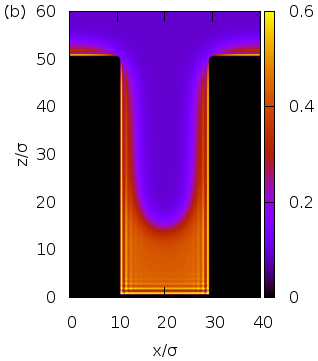}\\
\includegraphics[width=4cm]{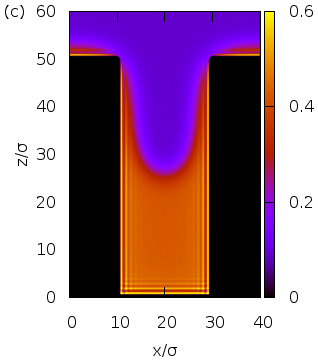} \includegraphics[width=4cm]{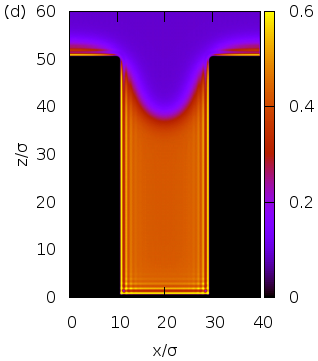}
\caption{Two-dimensional DFT equilibrium density profiles corresponding to condensation in the capillary groove of depth $D=50\,\sigma$ and width
$L=20\,\sigma$, at temperature $T=1.15\,T_w$. The departure $\delta\mu=\mu_{cc}-\mu$ from the chemical potential pertinent to capillary condensation
in an infinite slit of the same width is (in units of $\varepsilon$): a) 0.07, b) 0, c) -0.001, and d) -0.004.} \label{profs}
\end{figure}

Some more details of the fluid behaviour inside our capillary groove model can be revealed by determining two-dimensional density profiles as shown
in Fig.~\ref{profs} for several illustrative chemical potentials. We observe that in line with our expectations  a well pronounced meniscus forms at
the liquid-gas interface despite strong packing effects near the walls that make the fluid distribution strongly inhomogeneous. The meniscus indeed
shifts continuously upwards as the chemical potential increases from $\mu<\mu_{cc}$ to $\mu>\mu_{cc}$ and also note that thin wetting layers form at
the side walls as consistent with our slab model, unless the meniscus is located near the groove top.

\begin{figure}[h]
\includegraphics[width=8cm]{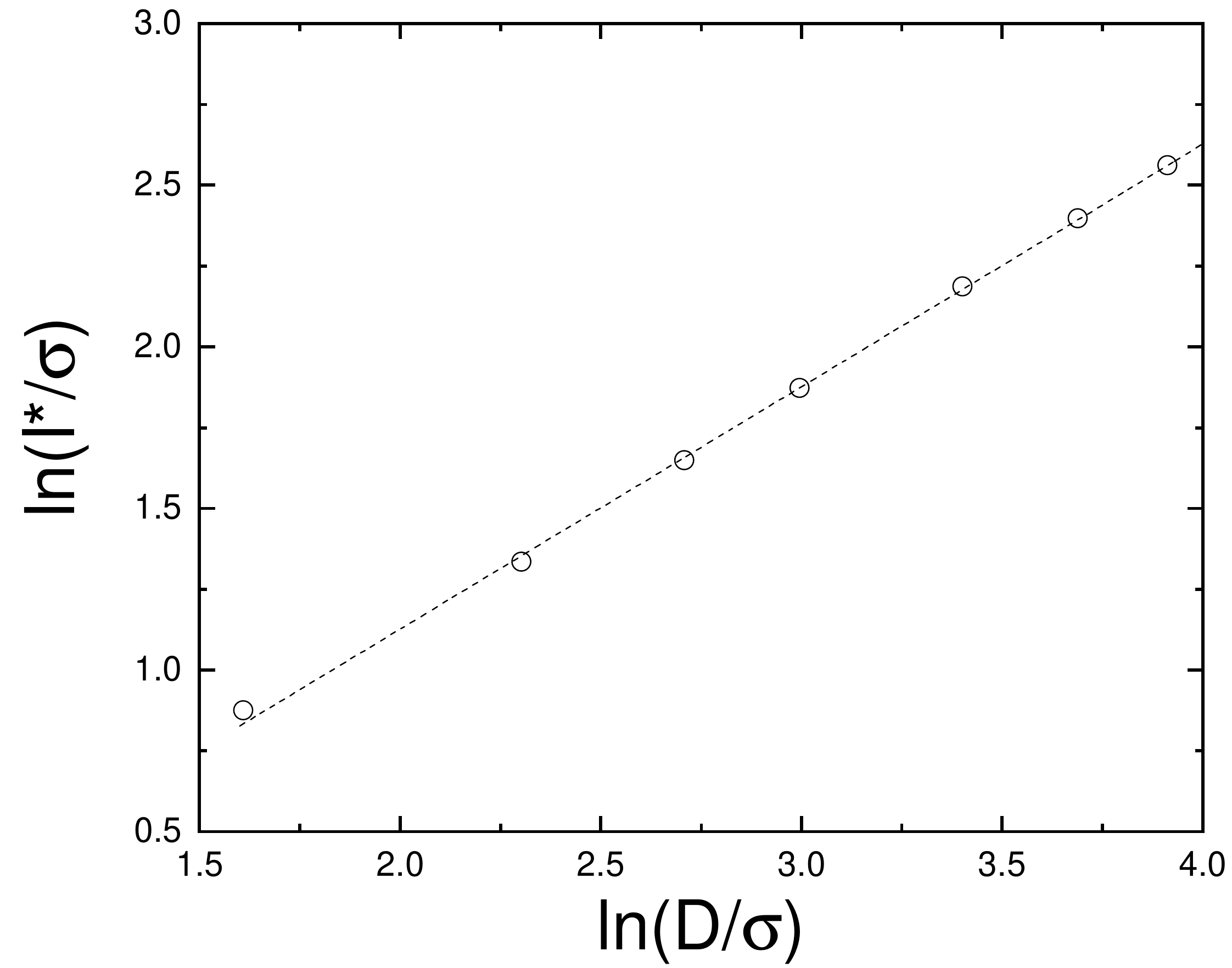}
\caption{A log-log plot of the dependence of the meniscus height $\ell^*=\ell(\mu_{cc})$ on groove depth $D$ as obtained from DFT (symbols). The
dashed line is the fit $y=y_0+0.75x$, a dependence corresponding to the slab model as given by Eq.~(\ref{ell_star}), to the DFT
data.}\label{fig_ell_star}
\end{figure}


Since grooves of finite $D$ are only partially filled with liquid at $\mu=\mu_{cc}$, one may enquire what  the dependence of the meniscus height
$\ell^*=\ell(\mu_{cc})$ is on the groove parameters. This height characterizes a location at which the condensation regime changes from $\ell\approx
\delta\mu^{-1/4}$ to $\ell\approx D-|\delta\mu|^{-1/3}$ and when the effective repulsive forces just balance. From the slab model it follows directly
that
  \bb
  \frac{\ell^*}{L}\sim\left(\frac{D}{L}\right)^{\frac{3}{4}}\,, \label{ell_star}
  \ee
for sufficiently large $D$. We test this result by comparing with the microscopic DFT model as is shown in Fig.~\ref{fig_ell_star} as a log-log plot;
we observe that the scaling form of Eq.~(\ref{ell_star}) is obeyed accurately even down to very small values of $D$ of the order of just units of
molecular diameters. This is surprising, sice one would expect any mesoscopic predictions to break down when $D<L/2$ not allowing for a meniscus
formation. It suggests that the nature of the condensation is governed by the presence of the long-range microscopic interactions (precisely included
in our slab model) rather than by a meniscus shape which was approximated rather crudely.

\begin{figure}[h]
\includegraphics[width=8cm]{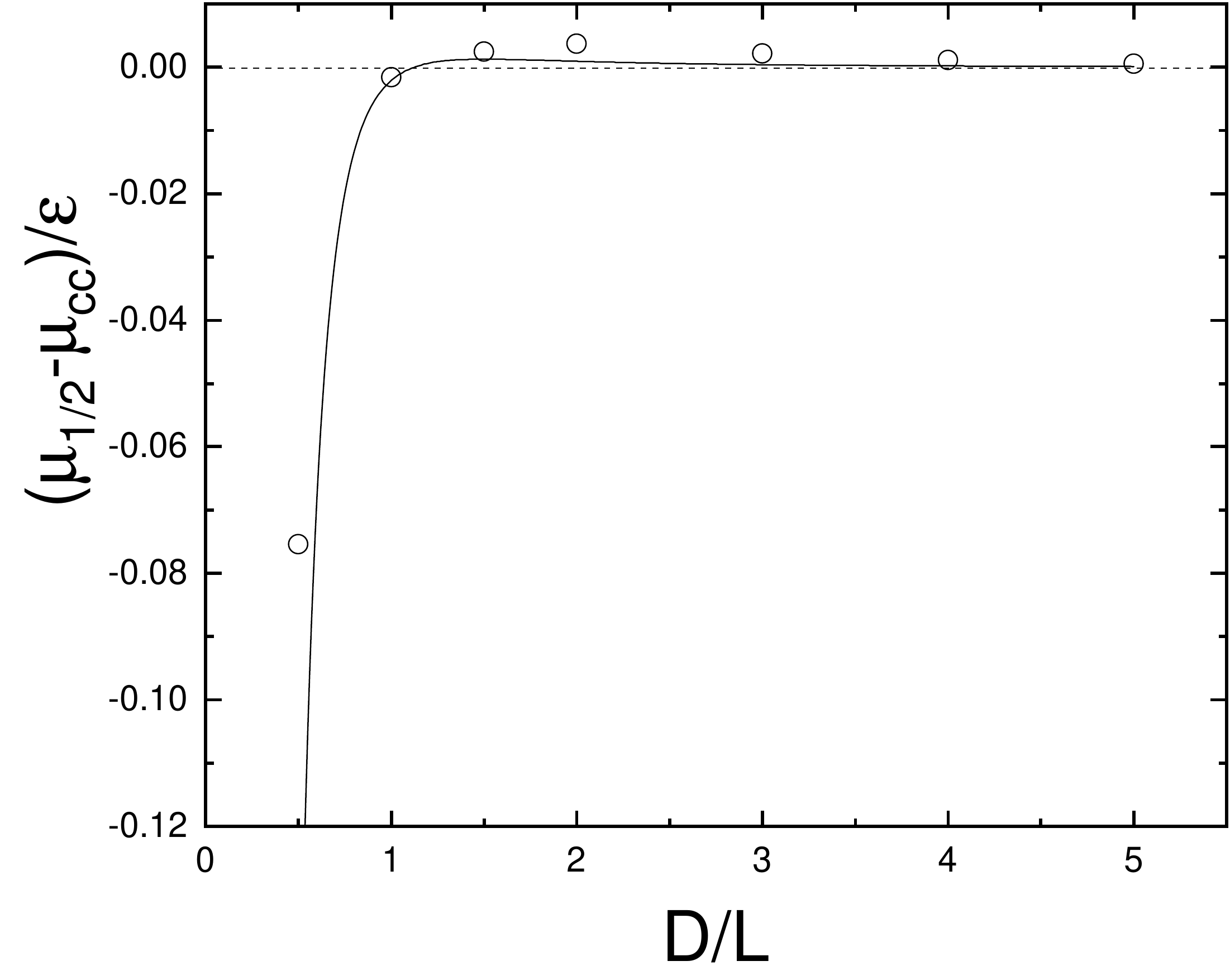}
 \caption{Dependence of the chemical potential $\mu_{\frac{1}{2}}$ at which the groove is half-filled with liquid, i.e. $\ell=D/2$, on the groove
 depth $D$. The symbols denote DFT results and the solid line denotes the prediction of the slab model, Eq.~(\ref{EL_slab}).}\label{fig_muhalf}
\end{figure}

The previous result, namely the non-linear dependence of $\ell(D)$, is a consequence of the asymmetry in the range of the two effective repulsions.
Another implication of this, also absent for infinitely deep grooves, can be addressed by determining the dependence of the chemical potential
$\mu_{\frac{1}{2}}$ at which the groove is half-filled with liquid on groove depth $D$. This is obtained by substituting for $\ell=D/2$ into
Eq.~(\ref{EL_slab}) which yields
 \bb
 \mu_{\frac{1}{2}}=\mu_{cc}+2a\left(\frac{8}{D^3}-\frac{9\,L}{D^4}\right)\,. \label{muhalf}
 \ee
From here it follows that $\mu_{\frac{1}{2}}(D)$ asymptotically approaches $\mu_{cc}$ from above as $1/D^3$. However, the function reaches its
maximum at $D=3L/2$ below which the last term  in (\ref{muhalf}) becomes dominating and $\mu_{\frac{1}{2}}$ decreases rapidly with decreasing $D$.
This also implies that there exists a unique finite value $D^*$ (for the given $L$ and $T$) for which the groove is half filled exactly at
$\mu_{cc}$. These predictions have been tested against the DFT model and the comparison is displayed in Fig.~\ref{fig_muhalf}. There is a good
overall agreement between the two theories although the location of the maximum of $\mu_{\frac{1}{2}}$ given by DFT is shifted to slightly larger
values of $D$. Nevertheless, both theories agree almost precisely that $D^*\approx L$, meaning that grooves of square cross-sections become half
filled with liquid at the pressure of the condensation of the capillary slits and macroscopic grooves.

\begin{figure}[h]
\includegraphics[width=8cm]{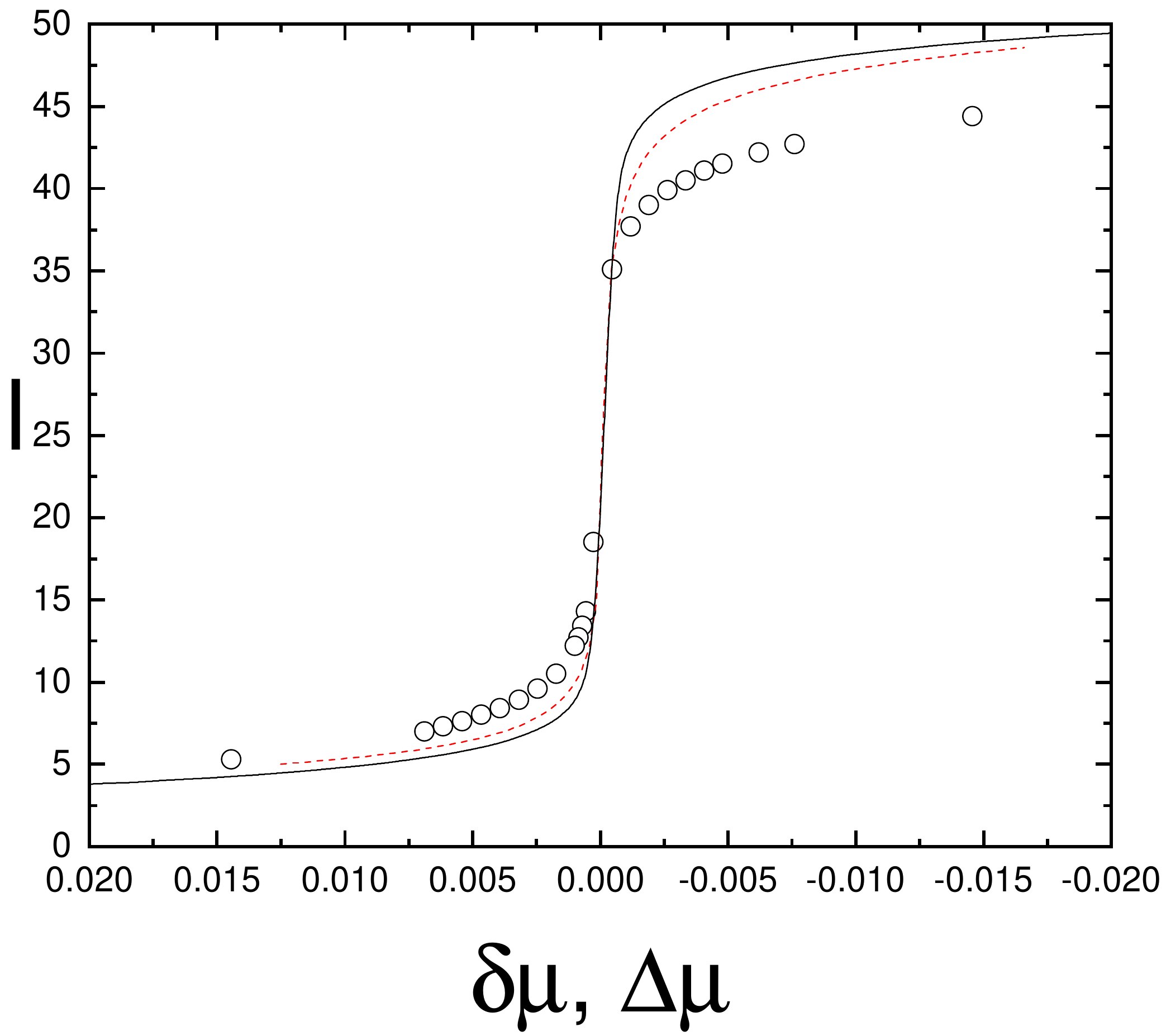}
\caption{Comparison of the location of the meniscus height $\ell(\delta\mu)$, where $\delta\mu=\mu_{cc}-\delta\mu$, in a capillary groove of depth
$D=50\,\sigma$ and width $L=20\,\sigma$ as obtained from microscopic DFT (symbols) and the full slab model using Eq.~(\ref{num_slab}) (red dashed
line). The black solid line shows the width of a wetting layer in a slit pore of width $D$ formed by two competing walls with external potentials
given by Eqs.~(\ref{potw}) and (\ref{potd}), as a function of $\Delta\mu=\mu_{\rm sat}-\mu$. Note that the abscissa is oriented such that $\mu$
increases from left to right.} \label{comp_walls}
\end{figure}

The previous results revealing the importance of the competition between the two effective interactions suggest that there exists a close analogy
between condensation in capillary grooves and in slit pores formed by a pair of plates with competing interactions. Within the latter model one
considers a strongly attractive wall favouring the liquid phase and a wall favouring the gas phase, a distance $D$ apart. Such a system may adopt
three stable configurations \cite{par_ev1, par_ev2}: a low-density state, when the slit is filled with capillary gas, a high-density state, when the
slit is filled with capillary liquid and a delocalized state, when the slit is partially filled with liquid and partially filled with gas. From a
mesoscopic viewpoint, the three states can be characterised by a liquid-gas interface which is either localized near one of the walls or delocalized
near a midpoint of the slit. In a bulk two-phase coexistence, $\mu=\mu_{\rm sat}$, a transition corresponding to the depinning of the interface from
either of the walls to the delocalized state may occur \cite{par_ev1, par_ev2}, the nature of which depends on the nature of the pertinent wetting
(or drying) transition. For the walls exhibiting first order transition, the localized-delocalized transition occurs at finite-size shifted wetting
temperature $T^*(D)$, such that $T_w<T^*<T_{sc}$ where $T_w$ is the wetting temperature and $T_{sc}$ is the prewetting critical temperature
\cite{mal_as}. Above $T_{sc}$, as is considered here, only the delocalized state is stable.

We now wish to map quantitatively the process of condensation in a groove of depth $D$ and width $L$ on the one in slit of width $D$ formed of
competing walls. Above $T_{sc}$ the latter corresponds to a continuous shift in the location of the liquid-gas interface from the vicinity of the
wetting wall to the drying one, as the chemical potential is varied around $\mu_{\rm sat}$. At the given $\mu$, the location of the interface
$0<\ell<D$ is determined by a balance of the effective and oppositely directed forces, induced by the long-range potentials of the walls. These
potentials, $V^{(w)}(z)$ and $V^{(d)}(z)$, are set  such that the excess grand potential per unit area for the slit model has the same structure as
the one for our groove model given by Eq.~(\ref{omegaex}):
 \bb
 \Omega^{\rm slit}=\Delta \mu\Delta\rho\ell+\frac{3A_w}{8\ell^3}+\frac{A_d}{(D-\ell)^2}\,.\label{slit_cw}
 \ee
except that the shift $\delta\mu=\mu_{cc}-\mu$ is replaced by $\Delta \mu=\mu_{\rm sat}-\mu$. Here,  $A_w\equiv
\pi\varepsilon_w^{(w)}\rho_w\sigma^6\Delta\rho/3$ and $A_d\equiv \pi\varepsilon_w^{(d)}\rho_w\sigma^6\Delta\rho/3$, with the parameters
$\varepsilon_w^{(w)}$ and $\varepsilon_w^{(d)}$  to be determined. Thus, while the first term on the rhs of Eq.~(\ref{slit_cw}) measures the free
energy penalty due to a departure from the two phase equilibrium, the last two terms are the binding potentials exerted by each wall. These are given
by \cite{dietrich}
 \bb
 W_\pi=-\Delta\rho\int_{\ell}^\infty V_\pi(z) \dd z\label{wpi}
 \ee
 where $V_\pi$ is the corresponding wall potential \cite{note}.
 From Eq.~(\ref{slit_cw}) it follows that the \emph{attractive} potential of the wetting wall is
 \bb
 V^{(w)}(z)=-\frac{\varepsilon_w^{(w)}\rho_w\sigma^6}{z^4}\,, \label{potw}
 \ee
 with $\varepsilon_w^{(w)}=3\pi L\varepsilon_w/8$ and the (long range) \emph{repulsive} potential of the drying wall is
 \bb
 V^{(d)}(z)=\varepsilon_w^{(d)}\rho_w\sigma^3\left(\frac{\sigma}{D-z}\right)^3 \label{potd}
 \ee
with $\varepsilon_w^{(d)}=\varepsilon_w$.

In Fig.~\ref{comp_walls} we display the DFT results of the meniscus growth in a groove of depth $D=50\,\sigma$ and width $L=20\,\sigma$ as a function
of $\delta\mu=\mu_{cc}-\mu$ and compare with the growth of the wetting layer in a slit pore with competing walls as a function of $\Delta\mu=\mu_{\rm
sat}-\mu$. The external long-ranged potentials of the slit walls given by Eqs.~(\ref{potw}) and (\ref{potd}) that largely determine the adsorption
behaviour in the slit are complemented with the rapidly decaying $\sim z^{-9}$ repulsive contributions that are kept the same as in Eq.~(\ref{vpi})
for both walls. We observe a reasonably good agreement showing a close link between the two processes especially for $\delta\mu>0$; for large values
of $\ell$ ($\delta\mu<0$) the deviation is slightly larger, which indicates that our slab model approximation becomes less accurate when the meniscus
reaches top of the groove;  this conclusion is indeed supported by inspection of the  density profiles shown in Fig.~\ref{profs}.

It should be emphasized that the simple structure of Eqs.~(\ref{EL_slab}) and (\ref{omegaex}) follows from the analysis of the slab model
(\ref{slab_model}) near $\mu_{cc}$ where the growth of the meniscus is most dramatic. Further away from $\mu_{cc}$, higher order terms neglected in
(\ref{omegaex}) may also become important. In this case, the last term in Eq.~(\ref{slab_model}) can be expressed as a single integral
 \begin{eqnarray}
&&\Delta\rho\int_{\tilde{\ell}}^D\dd z\int_{\ell_{\pi}}^{L-\ell_{\pi}}\dd x\,
V(x,z)\approx \nonumber\\
&&\approx2\alpha_6\Delta\rho\int_{\tilde{\ell}}^D\dd z
\left[\Psi(L-\ell_\pi,z)-\Psi(\ell_\pi,z)\right.\label{psi}\\&&\left.+\Psi(L-\ell_\pi,D-\tilde{\ell})-\Psi(\ell_\pi,D-z)
+\frac{L-2\ell_\pi}{z^3}\right]\nonumber
 \end{eqnarray}
where we only considered the attractive portion of $V(x,z)$ and introduced
 \bb
 \Psi(x,z)=\int \psi_6(x,z)\,\dd x=\frac{(2x^2-z^2)\sqrt{x^2+z^2}}{2x^2z^3}-\frac{x}{z^3}\,.
 \ee
The mean height of the meniscus $\ell(\mu)$ is given by minimization of (\ref{slab_model}),
$\left.\frac{\dd\omega^{ex}}{d\tilde{\ell}}\right|_{\tilde{\ell}=\ell}=0$, so that the integral in (\ref{psi}) does not need to be evaluated and we
obtain immediately
 \begin{eqnarray}
\mu&=&\mu_{\rm sat}-\frac{2\gamma}{\Delta\rho(L-2\ell_\pi)}\label{num_slab}\\
&&+\frac{2\pi\varepsilon_w\rho_w\sigma^6}{L-2\ell_\pi}\left[\Psi(L-\ell_\pi,\ell)-\Psi(\ell_\pi,\ell)\right.\nonumber\\
&&\left.+\Psi(L-\ell_\pi,D-\ell)-\Psi(\ell_\pi,D-\ell)+\frac{L-2\ell_\pi}{\ell^3}\right]\nonumber\,.
 \end{eqnarray}
Associating $\ell_\pi$ with the width of a wetting layer on an infinity planar wall we can write \cite{dietrich}
 \bb
\ell_\pi=\left(\frac{2\pi\varepsilon_w\rho_w\sigma^6}{3\Delta\mu}\right)^{\frac{1}{3}}
 \ee
and substituting for the values of the microscopic and thermodynamic parameters as used in our DFT, the location of the meniscus $\ell(\delta\mu)$
can be obtained from Eq.~(\ref{num_slab}) numerically. The resulting dependence is shown in Fig.~\ref{comp_walls} and we observe that the agreement
with DFT is better than for the slit with asymmetric walls, as expected, but the improvement is not dramatic. This implies that our slab model is
limited mainly by the approximation of the meniscus shape while neglecting the higher order terms in (\ref{omegaex}) is less significant.

\section{Conclusion}

In this work we studied condensation in capillary grooves of depth $D$ and width $L$ formed of completely wet walls interacting with a confined fluid
via long-range (dispersion) forces. As has been shown previously, condensation in macroscopically deep grooves, $D\to\infty$, is a critical process,
such that an amount of adsorbed liquid, which can be characterized by a meniscus height $\ell$, unbinds continuously from the groove bottom and
eventually diverges according to $\ell\sim\delta\mu^{-1/4}$, as $\delta\mu=\mu_{cc}-\mu$ tends to zero, in some analogy to complete wetting phase
transition on a planar wall. For $D$ finite, the transition becomes rounded but still experiences the same power-law behaviour for small (positive)
$\delta\mu$ as for infinitely deep grooves. However, this behaviour eventually breaks down in a very close neighborhood of $\mu_{cc}$, characterized
by the value $\bar{\delta}\mu$ which decays as $D^{-3}$ with the groove depth. For $\delta\mu$ negative, or indeed for
$\mu>\mu_{cc}+\bar{\delta}\mu$, the character of the condensation crosses over to the second regime where $\ell\sim D-|\delta\mu|^{-1/3}$. The
behaviour of the meniscus growth can be explained using a simple slab model from which it follows that the meniscus is effectively the subject of two
competing repulsive forces that act from the groove bottom and groove top, as a result of the presence of long-range intermolecular forces. The
trade-off between them has a number of further consequences. In particular, right at $\mu_{cc}$ the two repulsions are balanced out and the meniscus
is located at a height $\ell^*$ which scales as $(D^3L)^{1/4}$; note that the asymmetry in the range of the effective repulsions implies that the
relative filling of grooves at $\mu=\mu_{cc}$ decays with the aspect ratio $D/L$ as $\ell^*/D\sim (D/L)^{-1/4}$. Furthermore, we showed that the
chemical potential $\mu_{\frac{1}{2}}$ at which grooves are half-filled with liquid ($\ell(\mu_{\frac{1}{2}})=D/2$) exhibits non-monotonic dependence
on $D$, such that $\mu_{\frac{1}{2}}=\mu_{cc}$ for $D\to\infty$ and $D\approx L$, and drops rapidly well below $\mu_{cc}$ for $D<L$. Finally, we made
an explicit connection between condensation in capillary grooves and condensation in infinite slits made of asymmetric walls in a delocalized state.
One of the walls interacts with the fluid via retarded dispersion forces at long distances with the potential strength depending on $L$, while the
other wall, placed a distance $D$ apart, interacts with the fluid with a non-retarded dispersion potential but repulsively. It should be noted that
the model of the asymmetric slit where the density profile varies only in the direction perpendicular to the walls, is computationally much more
tractable than that of the capillary groove where the density profiles varies in two dimensions.

These predictions may have some interesting applications in modern technologies. With advanced techniques in nano-litography that enable the
modification of the shape of solid surfaces on molecular scales, the results suggest a simple mechanism of how to control an amount of adsorbed
liquid on the microscopic level. Consider a solid wall into which a network of capillary grooves is carved. From the slab model it follows that a
small change in the chemical potential from $\mu=\mu_{cc}-\delta\mu$ to $\mu=\mu_{cc}+\delta\mu$, with $\delta\mu>0$, induces the growth of the
meniscus height by the value $\delta\ell=D-(A/\delta\mu)^{1/3}-(AL/\delta\mu)^{1/4}$. This tells us that the adsorption responses sensitively by
tuning the chemical potential, i.e. the vapour pressure, around $\mu_{cc}$ ($p_{cc}$), and scales linearly with the groove depth $D$. Thus, even a
very small change in external conditions can be used to control the amount of the liquid adsorbed in a micro-porous medium which can still be
maintained (macroscopically) dry on its top and may thus be used as a storage of the adsorbate.

We conclude with some remarks regarding phenomena that have been omitted in this work and its possible extensions. Firstly, capillary grooves may
exhibit a pre-filling characterised by a jump in the meniscus height for temperatures near the wetting temperature \cite{mal_groove}; this is closely
related to pre-wetting but in contrast to the latter is one-dimensional in nature and thus becomes necessarily rounded if thermal fluctuations are
considered. The temperature considered here was deliberately chosen high enough to be beyond this temperature range, so that the condensation is
always continuous even on a mean-field level. Secondly, both treatments used in this work, the slab model and DFT, neglect the effect of interfacial
fluctuations, such as those in the meniscus height along the groove. However, the only effect of the fluctuations is that the condensation asymptotic
regime $\ell\sim\delta\mu^{-1/4}$ would eventually crossover to $\ell\sim\delta\mu^{-1/3}$ in a very close neighborhood of $\mu_{cc}$
\cite{our_groove}; this effect is utterly irrelevant for finite $D$ though since this region which is of the order of ${\cal{O}}(L^{-11})$ overlaps
with the crossover region of the characteristic width $\bar{\delta}\mu$ as given by Eq.~(\ref{bar_delta}), anyway. Possible extensions of the current
work include a study of a model of heterogeneous grooves where the side walls are of a different material than the bottom wall, extending
Ref.~\cite{het_groove} to finite values of $D$. Compared to the present study, this would not affect the nature of the effective repulsion from the
groove top which would still contribute as $\propto(D-\ell)^{-2}$ to the grand potential but the repulsion from the groove bottom would now be
$\propto \ell^{-2}$, i.e. the forces are of the same range. One of the consequences would be that $\ell^*$ in now independent of $L$ and scales
linearly with $D$, with the proportionality constant given by the difference in the wall potential strengths. Also, it would be interesting to
inspect the impact of the finite-size effects at different wall geometries, such as linear wedges or cylinders. Finally, the extension of the current
model to a periodic system of parallel grooves would allow to investigate the nature of the liquid-gas interface as $\mu\to\mu_{\rm sat}$ and the
effect of its undulation to the process of complete wetting. Some of these problems will be subject of our future work.

\section{Appendix A. Estimation of $\bar{\delta}\mu$}

For $\mu\approx\mu_{cc}$, the magnitudes of the effective repulsive terms in Eq.~(\ref{EL_slab}) are comparable while for a sufficient deviation from
$\mu_{cc}$ one of the two terms in the bracket becomes dominant; in this Appendix we estimate such a minimal deviation $\bar{\delta}\mu$ for each
case.

To this end, we define the new length-scale $x\equiv (a/|\bar{\delta} \mu|)^{1/3}$ and express Eq.~(\ref{EL_slab}) in the form
 \bb
 \frac{1}{x^3}\approx\frac{L}{\ell^4}-\frac{1}{(D-\ell)^3} \label{a1}
 \ee
where we ignored unimportant multiplicative constants. We now consider the cases $\mu<\mu_{cc}$ and $\mu>\mu_{cc}$ separately.

\subsection{$\mu<\mu_{cc}$}
In this case we are looking for a condition under which
 \bb
 \frac{L}{\ell^4}\gg\frac{1}{(D-\ell)^3} \label{a2}
 \ee
 implying
 \bb
 D\gg x+\ell \label{a3}
 \ee
 and
 \bb
 x\approx \left(\frac{\ell^4}{L}\right)^{\frac{1}{3}}\,. \label{a4}
 \ee
 By combining Eqs.~(\ref{a3}) and (\ref{a4}) it follows that
  \bb
  D\gg\ell\left[1+\left(\frac{\ell}{L}\right)^{\frac{1}{3}}\right]\,. \label{a5}
  \ee

We may now distinguish between two regimes, depending on the  relative values of $\ell$ and $L$. If $\ell\gg L$, then the last term in Eq.~(\ref{a5})
dominates compared to unity and thus $D\gg(\ell^4/L)^{\frac{1}{3}}$. Upon using  Eq.~(\ref{a4}) and substituting for $x$ this implies that the
relation in Eq.~(\ref{a2}) is obeyed provided $\bar{\delta}\mu\gg a/D^3$.

For $\ell\ll L$, the last term in Eq.~(\ref{a5}) is negligible compared to unity and the condition (\ref{a2}) requires that $D\gg\ell$ leading to
$\bar{\delta}\mu\gg aL/D^4$. This differs from the previous result only if $L$ appreciably deviates from $D$. Clearly, the possibility $L\gg D$ is
excluded since the groove geometry would not allow for a meniscus formation. On the other hand, $L\ll D$ would mean that the condition
$\bar{\delta}\mu\gg a/D^3$ is more restrictive than $\bar{\delta}\mu\gg aL/D^4$. This is however straightforward to show that within the former
regime $x\gg\ell\gg L$ whilst in the latter case we have $x\ll\ell\ll L$ and recalling that $\bar{\delta}\mu\propto 1/x$ these results contradict the
previous statement.

Finally, the intermediate case $\ell\sim L$ can only be realized if $L\ll D$ in order the condition (\ref{a2}) to be fulfilled. Then (\ref{a3})
reduces to $D\gg x$ leading again to $\bar{\delta}\mu\gg a/D^3$.

\subsection{$\mu>\mu_{cc}$}

Above $\mu_{cc}$ the situation is just opposite, i.e. we require
  \bb
 \frac{L}{\ell^4}\ll\frac{1}{(D-\ell)^3}\,, \label{a6}
 \ee
 from which (together with Eq.~(\ref{a1})) it follows that
 \bb
 D\approx x+\ell \label{a7}
 \ee
 and
 \bb
 D\ll\ell\left[1+\left(\frac{\ell}{L}\right)^{\frac{1}{3}}\right]\,, \label{a8}
 \ee
 Now, since $\ell<D$, we can write
 \bb
 D\ll\left(\frac{\ell^4}{L}\right)^{\frac{1}{3}}\,,
 \ee
 which on using of (\ref{a7}) leads after some rearrangements to
  \bb
 D\left[1-\left(\frac{L}{D}\right)^{\frac{1}{4}}\right]\gg x\,.
 \ee
 Since $x>0$, the last condition is satisfied only if $D\gg x$ reproducing (\ref{bar_delta}) again.

\begin{acknowledgments}
\noindent The support from the Czech Science Foundation, project 17-25100S, is acknowledged.
\end{acknowledgments}


\begin{thebibliography}{99}

\bibitem{row}
J. S. Rowlinson and B. Widom,  {\it Molecular Theory of Capillarity}, Oxford, Clarendon, (1982).

\bibitem{evans90}
R. Evans, J. Phys.: Cond. Matter. {\bf 2}, 8989 (1990).

\bibitem{der}
B. V. Derjaguin, Zh. Fiz. Khim. {\bf  137}, 14 (1940).

\bibitem{ev_mar85}
R. Evans and U. Marini Bettolo Marconi, Chem. Phys. Lett. {\bf 114}, 415 (1985).

\bibitem{ev_mar_tar86}
R. Evans, U. Marini Bettolo Marconi, and P. Tarazona, J. Chem. Soc. Faraday Trans. {\bf 2},  1763 (1986).

\bibitem{evans79}
R. Evans, Adv. Phys. {\bf 28}, 143 (1979).



















\bibitem{darbellay}
G. A. Darbellay and J. M. Yeomans, J. Phys. A {\bf 25}, 4275 (1992).

\bibitem{evans_cc}
C. Rasc\'on, A. O. Parry, N. B. Wilding, and R. Evans, Phys. Rev. Lett. {\bf 98}, 226101 (2007).

\bibitem{tasin}
M. Tasinkevych and S. Dietrich, Eur. Phys. J. {\bf E23}, 117 (2007).

\bibitem{mistura}
L. Bruschi and G. Mistura, J. Low Temp. Phys. {\bf 157}, 206 (2009).

\bibitem{hofmann}
T. Hofmann, M. Tasinkevych, A. Checco, E. Dobisz, S. Dietrich, and B. M. Ocko, Phys. Rev, Lett. {\bf 104}, 106102 (2010).

\bibitem{schoen}
H. Boelen, A. O. Parry, E. Diaz-Herrera and M. Schoen, Eur. Phys. J. E {\bf 25}, 103 (2008).

\bibitem{mal_groove}
A. Malijevsk\'y, J. Chem. Phys. {\bf 137}, 214704 (2012).

\bibitem{parry_groove}
C. Rasc\'on, A. O. Parry, R. N\"{u}rnberg, A. Pozzato, M. Tormen, L Bruschi, and G. Mistura, J. Phys.: Condens. Matter {\bf 25}, 192101 (2013).

\bibitem{ser}
P. Yatsyshin, N. Savva, and S. Kalliadasis, Phys. Rev. E {\bf 87}, 020402 (2013).

\bibitem{mal_13}
A. Malijevsk\'y, J. Phys.: Cond. Matter {\bf 25}, 445006 (2013).

\bibitem{mistura13}
G. Mistura, A. Pozzato, G. Grenci, L. Bruschi, and M. Tormen, Nat. Commun. {\bf 4}, 2966 (2013).

\bibitem{our_groove}
A. Malijevsk\'y and A. O. Parry,  J. Phys: Condens. Matter {\bf 26},  355003  (2014).

\bibitem{monson}
D. Schneider, R. Valiullin, and P.A. Monsosn, Langmuir {\bf 30}, 1290 (2014).

\bibitem{fan}
C. Fan, D.D. Do, and D. Nicholson, Mol. Simul. {\bf 41}, 245 (2014).

\bibitem{bruschi}
L. Bruschi, G. Mistura, P.T.M. Nguyen, D.D. Do, D. Nicholson, S. J. Park, and W. Lee, Nanoscale {\bf 7}, 2587 (2015).


\bibitem{fin_groove_prl}
A. Malijevsk\'y and A. Parry, Phys. Rev. Lett., in press.





\bibitem{ros}
Y. Rosenfeld,  Phys. Rev. Lett. {\bf 63}, 980 (1989).

\bibitem{dietrich}
S. Dietrich, in {\it Phase Transitions and Critical Phenomena}, edited by C. Domb and J. L. Lebowitz (Academic, New York, 1988), Vol. 12.




\bibitem{par_ev1}
A. O. Parry and R. Evans, Phys. Rev. Lett. {\bf 64}, 439 (1990).

\bibitem{par_ev2}
A. O. Parry and R. Evans, Physica A {\bf 181}, 250 (1992).

\bibitem{mal_as}
A. Malijevsk\'y, Cond. Matt. Phys. {\bf 19}, 13604 (2016).

\bibitem{het_groove}
A. O. Parry, A. Malijevsk\'y and C. Rasc\'on,  Phys. Rev. Lett. {\bf 113},  146101  (2014).

\bibitem{note}
We recall that the fluid-fluid potential is truncated, i.e. short-ranged, and thus does not contribute to the binding potential.










\end{thebibliography}
\end{document}